\documentclass[a4paper]{article}

\usepackage{INTERSPEECH2022}

\title{Automatic Dialect Density Estimation for African American English}
\name{Alexander Johnson$^1$, Kevin Everson$^2$, Vijay Ravi$^1$, Anissa Gladney$^3$, Mari Ostendorf$^2$, Abeer Alwan$^1$}
\address{
  $^1$UCLA, Department of Electrical and Computer Engineering\\
  $^2$University of Washington, Department of Electrical and Computer Engineering\\
  $^3$UCLA, Department of Linguistics}
\email{\{ajohnson49,vijaysumaravi,anissadg,alwan\}@ucla.edu, \{kpever,ostendor\}@uw.edu}

\begin{document}

\maketitle
\begin{abstract}
In this paper, we explore automatic prediction of dialect density of the African American English (AAE) dialect, where dialect density is defined as the percentage of words in an utterance that contain characteristics of the non-standard dialect. We investigate several acoustic and language modeling features, including the commonly used X-vector representation and ComParE feature set, in addition to information extracted from ASR transcripts of the audio files and prosodic information.  To address issues of limited labeled data, we use a weakly supervised model to
project prosodic and X-vector features into low-dimensional task-relevant representations.  An XGBoost model is then used to predict the speaker’s dialect density from these features and show which are most significant during inference.  We evaluate the utility of these features both alone and in combination for the given task.  This work, which does not rely on hand-labeled transcripts, is performed on audio segments from the CORAAL database.  We show a significant correlation between our predicted and ground truth dialect density measures for AAE speech in this database and propose this work as a tool for explaining and mitigating bias in speech technology.

\end{abstract}

\section{Introduction}

This paper\footnote{This work is supported in part by the NSF.} presents a novel framework for automatically assessing a speaker's dialect density. Dialect density is the degree to which
speech includes aspects of a dialect separate from the mainstream dialect of the language \cite{a1}.  A commonly used dialect density measure is the number of non-standard dialect tokens that appear in an utterance divided by the number of words in the utterance.  Previous literature \cite{b0} shows that high-dialect speakers often face more socioeconomic discrimination and bias in early education than their low-dialect counterparts.  The work in \cite{a2} demonstrates that the performance of ASR systems trained to recognize General American English (GAE) degrades when performing recognition on high dialect density African American English (AAE).  As dialectal changes are composed of both phonological and morphosyntactic aspects \cite{a3}, dialect-heavy speech can present challenges to both the acoustic model and the language model (LM) of an ASR system.  There is currently a lack of training data needed to train ASR systems to recognize AAE speech with the same fidelity they display for GAE speech, and no investigation on low-resource methods has been done in this area before.  As a result, bias against speakers of AAE and other dialects can propagate through spoken language systems, creating inequitable outcomes in the technology.  

To alleviate this issue, one could create an ASR system that first predicts the dialect density of its user given a small amount of speech and subsequently adjusts its model or model hyperparameters to best recognize speech of the target speaker's dialect.  This strategy has been applied with wide success to accented speech recognition \cite{a4,c1,a5}.  However, several differences exist between accent and dialect identification that make it difficult to directly apply previous work to dialect density prediction: First, accents are generally considered to be a binary characteristic.  That is, each speaker either does or does not have a given accent.  Dialect density is considered to be a continuum in which some speakers of a dialect possess more dialectal attributes than others.  As a result, previous work in accent-robust ASR typically only considers accent identification as a binary classification problem and does not lend itself well to identifying the degree of difference from mainstream speakers.  Second, most accent identification systems do not explicitly consider grammar and within-language diction as relevant to their task. However, dialects can be composed of a variety of changes in pronunciation, grammar, and diction which make them difficult to characterize by looking at any one feature of speech individually.  In addition, speakers may also consciously change the amount of dialect they use depending on social context (translanguaging or code switching) \cite{a6}, making it difficult to assign a consistent dialect density measure to a long portion of a speaker's speech as one would with a type of accent.  

For these reasons, new methods are needed in order for machines to accurately infer dialect density.  Although a primary use of this technology is in dialect-robust ASR, dialect density estimation could also be beneficial in data mining speech of a particular dialect, ensuring fair training of speech-language technologies, and evaluating human bias against speakers of dialects.  In this work, we propose a new method of automatic dialect density estimation for African American English from short utterances in a low-resource task.  We first analyze the correlation of several commonly-used and novel speech features in predicting dialect density (Sec. 2).  We then utilize a combination of the most effective of those features in order to create a model that best estimates the dialect density of a target speaker (Sec. 3). We follow our results with a discussion (Sec. 4) and conclusions (Sec. 5) on the usefulness of the features in predicting dialect density. 

\section{Methods}

\subsection{Data}
We use speech samples from the Corpus of Regional African American Language (CORAAL) \cite{a7}.  This database contains spoken interactions between an interviewer and an interviewee who speaks a regional variant of AAE.  The set of speakers range in age from approximately 20 to 80 years old and contains roughly equal numbers of male and female identifying participants.  The original sampling frequency is 44.1kHz. We down-sampled the audio files to 16kHz in all experiments.  We used the following numbers of speakers with regional AAE dialects from five US cities: 22 speakers from Washington DC (DCB), 10 speakers from Princeville, NC (PRV), 11 speakers from Rochester, NY (ROC), 10 speakers from Lower East Side Manhattan, NY (LES), and 12 speakers from Valdosta, GA (VLD).  For each speaker, several utterances with good audio quality ranging in length from 5sec to 1min were selected, and their dialect densities were scored by hand as ground truths.  The subset of speakers from DCB, PRV, and ROC were chosen because their dialect densities were annotated by the authors of \cite{a2}. All speakers from VLD and LES were used, and their dialect densities were scored by the authors of this paper.  This results in a total of approximately 3 hours of dialect density-scored utterances from 65 speakers.  This data set was split into speaker independent sets for which 70\% was used in training, 15\% was used in validation, and 15\% was used in testing.

\subsection{Feature Extraction}

From each utterance, we extracted the following six feature sets:

\textbf{ Wav2Vec2.0 Transcripts}: For the first three feature sets, we generated ASR transcripts using a pretrained Wav2Vec2.0 model \cite{a8} trained on the 960hr LibriSpeech database \cite{a9}.  While these transcripts contained errors and misrepresented the out-of vocabulary (OOV) words, we implicitly attempted to utilize consistent errors and accurate portions of the transcripts to identify useful phonetic and grammatical information.

\textbf{1. ASR Output Character Combination Frequency}: The frequency of each sequence of two characters (bigram) in the transcript was counted and used as a feature.  The Wav2Vec2.0 model can output 31 different characters leading this feature to be a 961 x 1 vector which can be thought of as the flattened 31 x 31 matrix in which the element in row i and column j is the number of times character i was followed directly by character j in the generated transcript for the given utterance.  We hypothesize that this feature will capture consonant clusters that commonly occur in a particular dialect.

\textbf{2. ASR Output Character Duration}: From the output logits of the Wav2Vec2.0 model, the average duration of each output character was computed.  We hypothesize that this feature will be useful in determining which sounds are more or less frequently spoken or stressed by speakers of a particular dialect.

\textbf{3. ASR Output Language Modeling}: In addition to the previously mentioned features, we were interested in how neural language modeling techniques could be applied to automatically generated transcripts of speech in order to predict dialect density.  We noticed that, of the most commonly noted features of AAE \cite{a1,a2}, language differences relating to the tense, collocation, and negation of verbs (eg. absence of copula, negative concord, generalization of ``is" and ``was" to use with plural and second person subjects, etc.) were especially prevalent.  This led us to pay particular attention to verbs. First, the verbs in each utterance were found using a pre-trained FLAIR part-of-speech tagging model \cite{a18, a19}. We then used the Fisher corpus \cite{a10}, consisting largely of GAE conversations, to train word- and character-based LSTM language models, which provide probability distributions over the next word or character in an utterance given the history. To measure mismatch of verbs in the GAE training data and AAE testing data, we then extracted the verb OOV rate (using the word-based model vocabulary) and the average verb surprisal \cite{b2} (using the character-based model) for each utterance, where the surprisal of the $i$-th word ($S(w_i)$) is calculated from the letter LM as:
\begin{eqnarray*}
S(w_i) & =& -\log(p(w_i|w_{i-1},\ldots ,w_1))\\
 & = & -\sum_{j=1}^{l_i}\log(p(c_{i,j}|c_{i,j-1}, \ldots ,  c_{1,1}))
 \end{eqnarray*}
%
%
where $l_i$ is the length of the word $w_i$ with characters $c_i,_j$.  We also calculate the overall utterance perplexity from the character-based model (char\_ppl), the average surprisal for all words, and the ratio of average verb surprisal to average overall surprisal. Since the LM is trained on GAE, word choices more characteristic of AAE will have high surprisal.

\textbf{4. CompareE16 Features}: The widely used ComParE16 features \cite{a12} were extracted from the audio segments using the OpenSmile toolkit \cite{a13}.  This set includes pitch, energy, spectral, cepstral coefficients (MFCCs) and voicing related frame-level features which are referred to as low-level descriptors (LLDs). It also includes the zero crossing rate, jitter, shimmer, the harmonic-to-noise ratio (HNR), spectral harmonicity and psychoacoustic spectral sharpness. In total, this feature set contains 6373 features resulting from the computation of various statistics, polynomial regression coefficients, and transformations calculated over the low-level descriptor contours.

\textbf{Weak Supervision}: To create the following two feature sets, we employed a weakly-supervised learning technique.  We noticed that the five cities used from the CORAAL database have widely varying average dialect
densities, with the averages from PRV and VLD being much higher than those from ROC and LES, and with the DCB average in between.  Therefore, we believed that an utterance's city of origin could serve as a weak label in a preliminary step before dialect density estimation.  We gathered the set of utterances from the entirety of the 200hr CORAAL database from the five cities of interest that matched the following criteria: 1) Contained at least 10 words to have enough speech to estimate dialect density, 
2) Contained no interruptions from the interviewer 3) Were not contained in the set of dialect density-scored utterances.  We then used shallow neural networks to map larger input feature vectors into 5-dimensional vectors for which the ith element represents the probability that the utterance was spoken by a speaker from the ith city in the database.  This step is intended to project larger sources of information into smaller features vectors which contain only relevant dialect information. The idea is that training a model  to classify diverse utterances by region would prompt it to learn region-specific information such as dialectal traits without the need to label the dialect density of all of the utterances in the training set.  The output 5-dimensional vector is then used as the representative feature.

\textbf{5. X-Vector}: The popular X-vector was incorporated to capture speaker-specific information \cite{a11}.  These 512-dimensional neural network-generated embeddings contain speaker-specific information that may relate to dialect.  As described earlier, the 512-dimensional vectors were projected into 5-dimensional feature vectors using the fully connected network shown in Figure 1.  This network achieved a validation accuracy of 72.6\%.

\textbf{6. Prosodic Embedding}: Inspired by \cite{a14}, four pitch and energy features were extracted across time from the utterances: F0 (extracted with Praat \cite{b6}), the total energy in the frame, the energy in the spectrum below 1kHz, and the energy in the spectrum above 1kHz.  These features were then normalized and used as the input to a CNN (as shown in Figure 1) that was trained to predict the region of origin of the speaker.  This forces the CNN to classify region specific information from only the prosodic information contained in the speaker’s changes in pitch and energy.  This CNN achieved a validation accuracy of 70.7\%.  The output probability vector was then used as the final prosodic embedding.

\begin{figure}[h]
\centering
\includegraphics[width=0.4\textwidth, height=6cm]{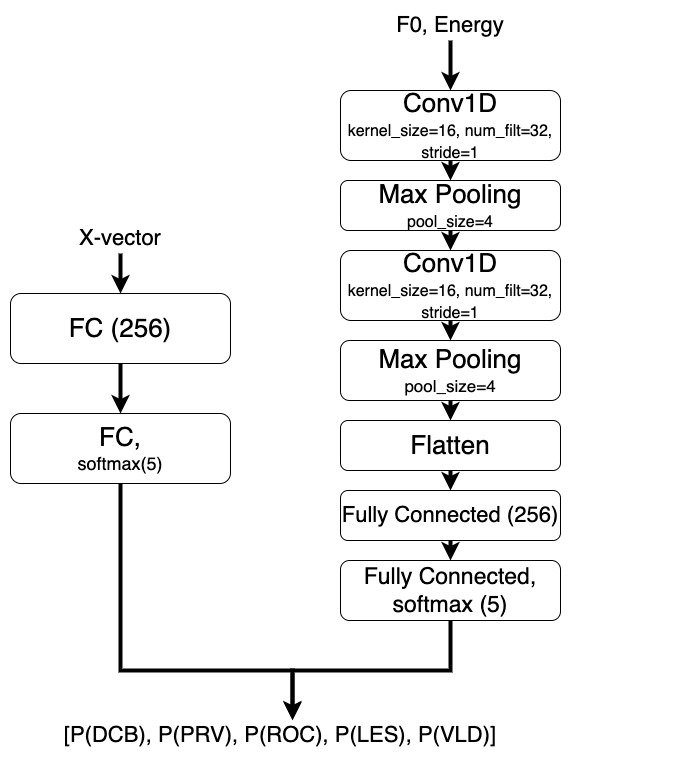}
\caption{The architecture for the fully connected (FC) network used to project the X-vectors (left) and the CNN used to project the prosodic information (right).  The inputs to the CNN are the pitch (F0) and three energy contours of the utterance. The output of both networks is a vector whose elements represent the probability of the speaker belonging to each of the cities used from the CORAAL database.}
\end{figure}

\begin{figure}[h]
\centering
\includegraphics[width=0.5\textwidth]{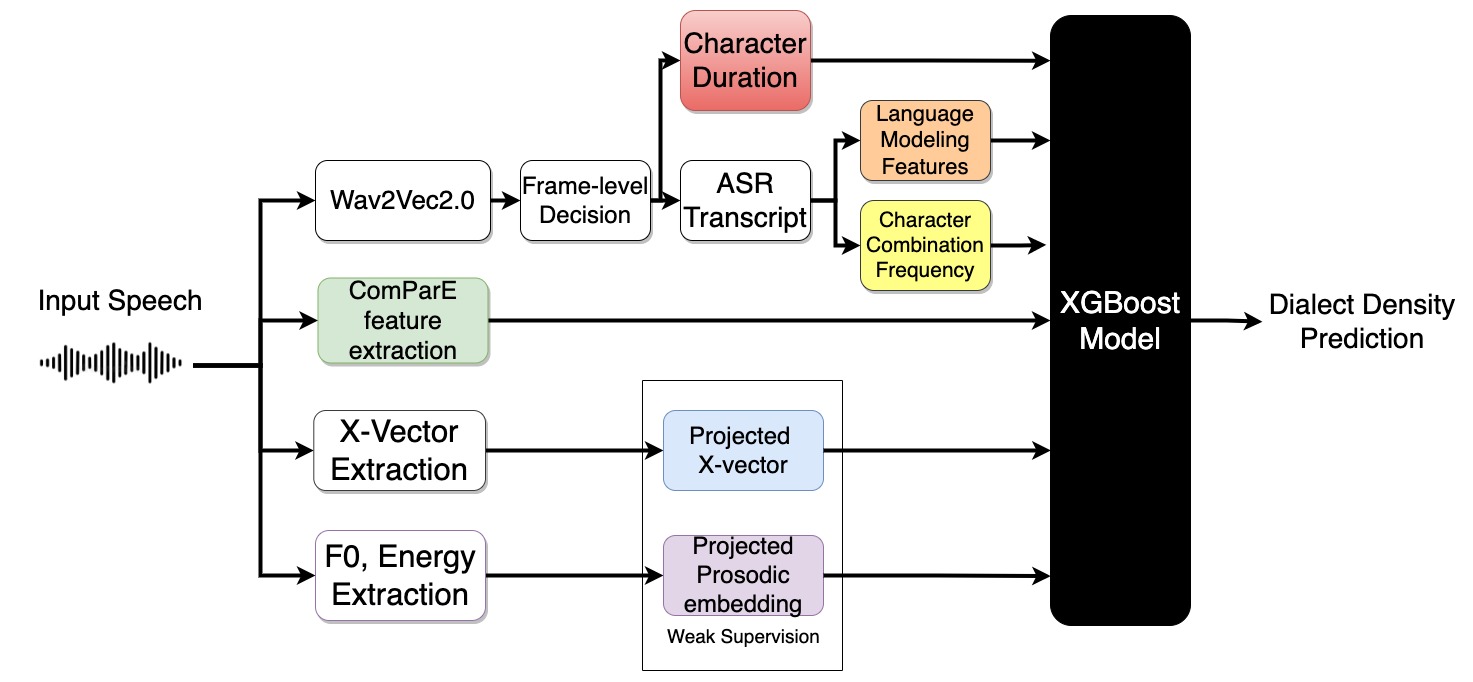}
\caption{Overview of the features used in the proposed work.}
\end{figure}
\subsection{Model Training}
First, one distinct XGBoost model \cite{a15} was trained for each of the six feature sets.  This boosted decision tree model has the advantage of allowing us to easily measure the impact of the input features on the output value for explainability.  Each of the six models was trained to predict dialect density scores from one of the given input feature sets.  Then the correlation between the predicted dialect density labels and actual dialect density labels was calculated.  We chose correlation as the performance metric because human-performed dialect density assessments are subject to possibly high inter-rater variability within the ranges of their scores \cite{a16}, and so evaluation methods that rely heavily on the absolute value of the dialect density may be subject to measurement noise.  However, raters do tend to assign higher or lower scores to the same speakers, and so we expect correlation between predicted and ground truth scores to be meaningful.  As some features may only correlate with phonological aspects or only correlate with morphosyntactic aspects of dialect density, we train each model to predict each of the three types of dialect density scores:
$$
DDM phon = \frac{N_{ph}}{N} \qquad DDM gram=\frac{N_{ms}}{N}
$$
$$
DDM=\frac{N_{ph} + N_{ms}}{N}
$$
where $N_{ph}$ is the number of phonological AAE tokens in the utterance, $N_{ms}$ is the number of morphosyntactic AAE tokens in the utterance, and \textit{N} is the total number of words in the utterance.  The average dialect density for each city used is shown in Table 1.

\begin{table}[h]
\centering
\begin{tabular}{ c c c c }
\hline
 & DDMphon & DDMgram & DDM\\
\hline
DCB & 0.083 & 0.004 & 0.088 \\
ROC & 0.041 & 0.006 & 0.047 \\
PRV & 0.166 & 0.028 & 0.194 \\
LES & 0.018 & 0.025 & 0.042\\
VLD & 0.122 & 0.029 & 0.141\\
\hline
\end{tabular}
\caption{\label{tab:table-name} Average dialect density by city for each of the dialect density measures shown.}
\end{table}


Finally, we used the set of all features as the input to the XGBoost model, as shown in Figure 2.  As the ComParE feature set was large, only the most impactful 10 ComParE features were used in the combined feature set.  

\section{Results}

Table 2 gives the Pearson Correlation of the predicted dialect density measure with the ground truth labels for the test set for an XGBoost model trained on the listed feature sets.
We also include the SHAP value plots \cite{a17} which give the relative importance of each feature to the model during prediction.  Figures 2 and 3 give the SHAP value plots for the models trained on all features for predicting DDMphon and DDMgram, respectively.  As the DDMphon term dominiates the total dialect density measure, the SHAP value plot for DDM is nearly identical to that of DDMphon.  In these plots, the Wav2Vec2.0 Char Comb features are listed as $char1\_char2$ (e.g. N\_space is the frequency of the ``N" character being followed by a space character), and the Wav2Vec2.0 Char Dur features are listed as the character whose duration was used as the input feature from the letters A-Z, period, apostrophe, space, or silence (sil) characters.

In order to demonstrate the reliability of our results, we also perform random hold out on the highest performing features.  Here, we randomly select speaker-independent train and test split (80\% train, 20\% test) from the data 200 times and report the average scores over all runs in Table 3.
\begin{table}
\begin{center}
\begin{tabular}{ |c|c|c|c| }
\hline
Correlation & DDMphon & DDMgram & DDM\\
\hline
\begin{tabular}{@{}c@{}}Wav2Vec2.0 \\ Char Dur.\end{tabular} & 0.382 & -0.013 & 0.359\\
\hline
\begin{tabular}{@{}c@{}}Wav2Vec2.0 \\ Char Comb\end{tabular} & 0.303 & 0.124 & 0.503\\
\hline
Wav2Vec2.0 LM & 0.520 & 0.108 & 0.637\\
\hline
X-vector & 0.404 & 0.392 & 0.369\\
\hline
ComParE & 0.102 & 0.189 & 0.443\\
\hline
Prosody & 0.029 & 0.376 & 0.008\\
\hline
All features & 0.552 & 0.430 & 0.718\\
\hline
\end{tabular}
\end{center}
\caption{\label{tab:table-name} Pearson Correlation between actual and predicted dialect density measures for each of the three metrics: Only the phonological component of the dialect density (DDMphon), only the morphosyntactic component of the dialect density measure (DDMgram), and the entire dialect density measure (DDM).  The results for the model trained on six feature sets individually as well as the model trained on the combination of all of the features are shown.}
\end{table}

\begin{table}
\begin{center}
\begin{tabular}{ |c|c|c|c| }
\hline
Correlation & DDMphon & DDMgram & DDM\\
\hline
\begin{tabular}{@{}c@{}}Wav2Vec2.0 \\ Char Comb\end{tabular} & 0.339 & 0.126 & 0.495\\
\hline
Wav2Vec2.0 LM & 0.502 & 0.173 & 0.629\\
\hline

All features & 0.569 & 0.385 & 0.678\\
\hline
\end{tabular}
\end{center}
\caption{\label{tab:table-name} Average Pearson Correlation between actual and predicted dialect density measures for each of the three DDMs over 200 iterations of Random Hold Out validation.}
\end{table}

\begin{figure}[h]
\centering
\includegraphics[width=0.5\textwidth,height=6cm]{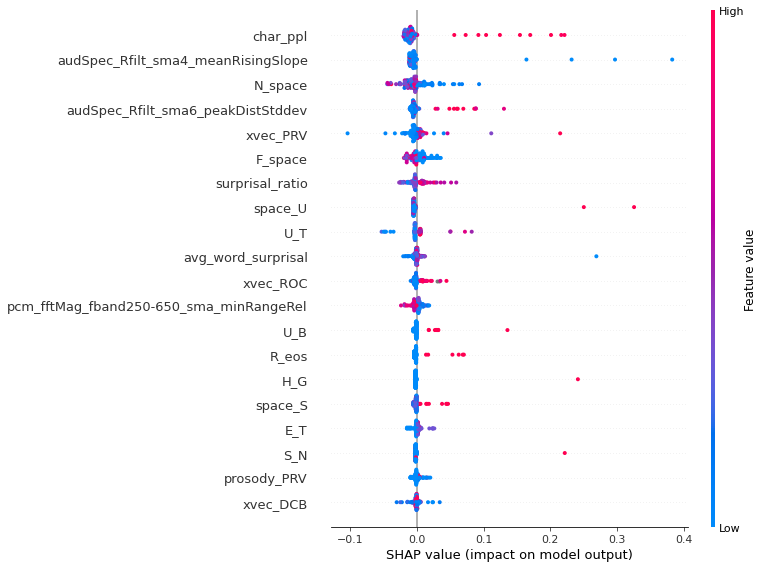}
\caption{SHAP value plot for the XGBoost model trained to predict DDMphon from the set of all features.  The features are listed from top to bottom in order of significance.}
\end{figure}

\begin{figure}[h]
\centering
\includegraphics[width=0.5\textwidth,height=6cm]{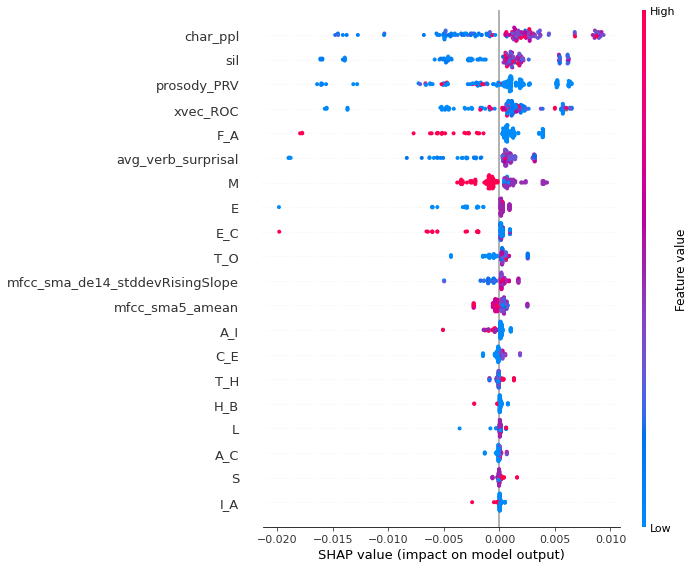}
\caption{SHAP value plot for the XGBoost model trained to predict DDMgram from the set of all features.  The features are listed from top to bottom in order of significance.}
\end{figure}

\section{Discussion}
Looking at the individual features, we note that the Wav2Vec2.0 Character Combinations and Wav2Vec2.0 LM features were especially effective in estimating dialect density.  Many of the character combinations appear to relate to word initial and word final sounds (eg. N\_space (frequency of an ``N" followed by a space character in the ASR transcripts), F\_space (frequency of an ``F" followed by a space), and space\_U (frequency of a space followed by a ``U")).  This is in line with observations that AAE includes dropping of word final nasals and glides and simplification of word initial and word final consonant clusters.  Character perplexity (char\_ppl) from the language modeling features was the most impactful feature in estimating all three DDM scores.  This feature is particularly useful in providing an objective distance metric between the GAE of the Fisher Corpus and the ASR transcripts of the target dialect speech which, unlike WER, does not require ground truth transcripts or suffer as heavily in the presence of OOV words.  The features derived through weakly supervised embedding (projected X-vector and Prosody embedding) have the most significant correlation with DDMgram.  This may indicate that learning grammar from audio files or imperfect transcripts requires larger amounts of data which our method of weak supervision allows us to utilize.  In general, the ComParE features using Auditory Rasta filtering proved to be most useful. The RASTA-style filtered auditory spectrum is inspired by psychoacoustics and has been shown to capture context-dependent information useful in speech recognition \cite{a20}.   

As Figure 3 shows, the combination of the five most impactful features in predicting DDMphon was: character perplexity (char\_ppl), mean rising slope of the Rasta-filtered auditory spectrum, the frequency of an ``N" character followed by a space character in the ASR transcripts (N\_space), the standard deviation of distances between peaks in the Rasta-filtered auditory spectrum, and the PRV component of the projected X-vector.  As Figure 4 shows, the five most impactful features in estimating DDMgram are character perplexity, duration of sounds predicted to be silence or unintelligible by Wav2Vec2.0 (sil), the PRV component of the prosody embedding, the ROC component of the projected X-vector, and the frequency of an ``F" followed by an ``A" in the ASR transcripts (F\_A).  The frequency of F\_A as a feature may due to a raising of the vowel following ``F" in several words such as ``fell" or ``fire" as is seen in some dialects of the US South.

\section{Conclusions}
In this work, we are able to predict dialect density with a high correlation 
($~68\%$) between the predicted and groundtruth labels as well as explain a possible combination of factors that make up dialectal differences in ASR. The results may improve when other factors are considered such as finer aspects of prosody or within-speaker variability.  The system may also improve with more accurate ASR transcripts.  Overall, we have created a novel framework from which to later explore these factors.  Future work includes further exploring the role of prosody in dialect, applying this framework to other dialects, and applying this work in dialect-robust ASR systems. 



\bibliographystyle{IEEEtran}


\end{document}